\documentclass[aps,prb,reprint,showpacs,showkeys]{revtex4-2}
\usepackage{graphicx} 
\usepackage{amsmath} 
\usepackage[version=4]{mhchem} 
\usepackage{bm} 
\usepackage{natbib}
\usepackage{natmove}
\usepackage[colorlinks=true, linkcolor=blue, citecolor=blue, urlcolor=blue, linktoc=page, bookmarks=false, pdfstartview={FitH}, pdfborder={0 0 0.0 [3 3]}]{hyperref} 
\usepackage{color} 
\usepackage{dcolumn} 
\newcolumntype{.}{D{.}{.}{2.5}}
\newcolumntype{-}{D{.}{.}{4.0}}
\usepackage{makecell}
\usepackage{multirow}
\usepackage{cleveref}
\usepackage{easyReview} 
\crefname{figure}{Fig.}{Figs}
\crefname{table}{Table}{Tables}
\renewcommand{\today}{\number\day \space \ifcase \month \or January\or February\or March\or April\or May\or June\or July\or August\or September\or October\or November\or December\fi \space \number\year} 
\def\m1r{\multicolumn{1}{r}}
\newcolumntype{P}[1]{>{\centering\arraybackslash}p{#1}}
\begin{document}
\title{Implications of electron and hole doping on the magnetic properties of spin-orbit entangled \ce{Ca4IrO6} from DFT calculations}
\author{Avishek \surname{Singh}}
\author{Jayita \surname{Chakraborty}}
\author{Nirmal \surname{Ganguli}}
\email[Electronic address: ]{NGanguli@iiserb.ac.in}
\affiliation{Department of Physics, Indian Institute of Science Education and Research Bhopal, Bhauri, Bhopal 462066, India}
\date{\today}
\begin{abstract}
	We investigate the electronic structure and magnetic properties of a $J_\text{eff} = 1/2$ iridate \ce{Ca4IrO6} and the implications of doping electrons and holes using {\em ab initio} density functional theory. Our calculations considering spin-orbit interaction reveal that although the Mott-insulating parent compound transforms into a conductor upon doping, antiferromagnetism sustains in the doped system, albeit with a grossly noncollinear arrangement of the spins. We find a strong spin-orbit interaction and magneto-crystalline anisotropy, causing frustration in the system, possibly leading to the highly noncollinear arrangement of spins upon non-magnetic doping. Our results may be important from the viewpoint of spintronics using iridates or other $5d$ materials.
\end{abstract}
\pacs{} 
\keywords{Iridate, Spin-orbit interaction, Noncollinear antiferromagnetism}
\maketitle
\section{\label{sec:intro}Introduction}
The effect of spin-orbit interaction (SOI) on $5d$ oxide materials has been demonstrated to be the origin of a number of exotic physical properties, including spin liquids \cite{okamoto2007}, $J_\text{eff} = 1/2$ Mott insulating state \cite{kim2008,Kim2009}, unconventional magnetism, and topological phases \cite{shitade2009}. In these materials, especially in iridium oxides, the energy of SOI, electron-electron correlation, and crystal field splitting are of comparable strength. Different spin-orbit magnetic states are realized in iridates depending on the oxidation states of Ir ions. In the presence of spin-orbit interaction, the $t_{2g}$ states of Ir$^{4+}$ ion in IrO$_6$ octahedra split into $J_\text{eff} = 1/2$ and $J_\text{eff} = 3/2$ multiplets. Being lower in energy, the $J_\text{eff} = 3/2$ band is completely filled, while the higher-energy $J_\text{eff} = 1/2$  band is half-filled. The half-filled $J_\text{eff} = 1/2$ band further splits into a completely occupied lower Hubbard band and an empty upper Hubbard band with the inclusion of a small Hubbard $U$ value in a theoretical model. While the compounds with Ir$^{4+}$ oxidation state (5$d^5$ electronic configuration) prefer a $J_\text{eff} = 1/2$ Mott-insulating state, a nonmagnetic singlet ground state is expected in pentavalent (5$d^4$) iridates \cite{IrScience}. On the other hand, spin-orbit driven magnetism with strong covalence with oxygen is found in $d^3$, and $d^{3.5}$ based transition-metal oxides \cite{ChakrabortyPRB18}. Distortion of the IrO$_6$ octahedra plays a critical role in determining the electronic structure of these compounds \cite{Phelan2015}.

Experimental studies on electron-doped iridate have shown persisting long-range antiferromagnetic interactions of undoped systems up to 4\% doping concentration in Sr$_{2-x}$La$_x$IrO$_4$. Beyond that, only short-range antiferromagnetic interactions survive \cite{cheng2015}. However, according to the Nagaoka theorem, a minimal amount of carrier doping can destroy the antiferromagnetic state within a nearest neighbor model \cite{nagaoka1966}. However, an analysis of the stability of the antiferromagnetic state for electron-doped iridates within Hartree-Fock approximation and second-order perturbation theory reveals the persistence of antiferromagnetic states for small doping concentration \cite{Bhowal2018}. With 1\% doping of carriers, \ce{Sr2IrO4} is found to exhibit a metallic state, suggesting a decrease in energy gap with increasing carrier concentration \cite{Han2016}. Further, increasing carrier concentration results in canted magnetic states, long-range exchange interactions, and unconventional superconductivity. A significant decrease in electrical resistivity along with a substantial change in magnetic and transport properties have been observed for 10\% hole and electron doping in \ce{Ca5Ir3O12} \cite{Haneta2019}. New electronic and magnetic orderings emerge with electron and hole doping in bilayer iridates due to fluctuation in local electron density that promotes nano-scale phase separation \cite{Wang}, leading to two different types of order coexisting within one $J_\text{eff} = 1/2$ band.

Fractional oxidation states of $5d$ transition metals in various compounds exhibit interesting properties like pressure-induced magnetic transition \cite{panda2015} and mixed valence dimers \cite{Dey2017}. A study of magnetism in \ce{Ba3YIr2O9} with Ir-$d^{4.5}$ configuration revealed strong inter-dimer coupling. Our previous work on \ce{LaAlO3}$|$\ce{SrIrO3}$|$\ce{SrTiO3} heterostructures with Ir-$d^{5.5}$ configuration revealed a noncollinear antiferromagnetic two-dimensional conducting system with Rashba and Dresselhaus spin-orbit interaction \cite{ChakrabortyPRB20}. Although ultrathin \ce{SrIrO3} film behaves as an antiferromagnetic Mott-insulator with a $J_\text{eff} = 1/2$ state, the observation of canted antiferromagnetic arrangement in monolayer \ce{SrIrO3} sandwiched between \ce{LaAlO3} and \ce{SrTiO3} hosting a two-dimensional conducting system was intriguing. Nagaoka physics suggests the antiferromagnetic state to transform into a ferromagnetic one upon carrier doping that leads to a conducting state.

The above discussion highlights the importance of investigating the interesting physical properties of mixed-valent iridates including electronic structure, conducting nature, magnetic order, spin-orbit interaction, and the noncollinear antiferromagnetism observed in $J_\text{eff} = 1/2$ antiferromagnetic Mott insulators doped with carriers. Doped \ce{Ca4IrO6} can work as a useful platform for this purpose. With a nominal Ir$^{4+}$ oxidation state, \ce{Ca4IrO6} is expected to exhibit a $J_\text{eff} = 1/2$ insulating state that has previously been theoretically investigated in other compounds \cite{Sarkozy1974, segal1996}. Here we investigate the electronic structure and magnetic properties of \ce{Ca4IrO6} and $J_\text{eff} = 1/2$ Mott insulator states via first-principles density functional theory calculations. Subsequently, we examine the effect of electron and hole doping on electronic and magnetic properties of \ce{Ca4IrO6}. The remainder of this article has been organized as follows. \Cref{sec:method} details the methodology adopted for the calculations performed. The results obtained from our calculations have been discussed in \cref{sec:result}. Finally, we summarize the work in \cref{sec:conc}.

\section{\label{sec:method}Crystal structure and method of calculations}
\begin{figure}
     \includegraphics[scale=0.8]{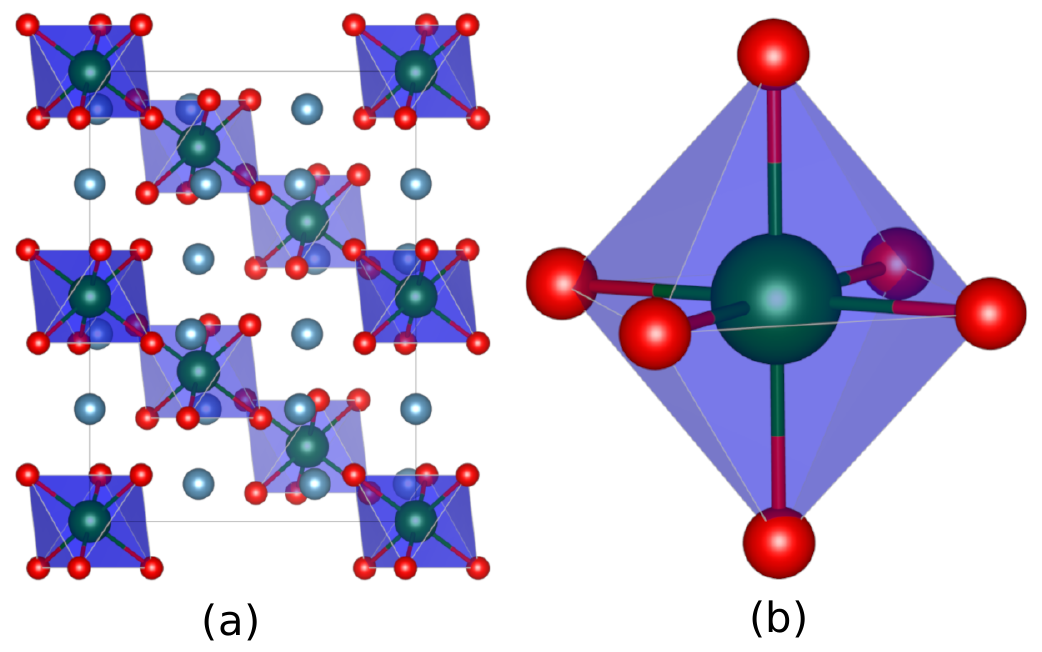}
    \caption{(a) The unit cell of Ca$_4$IrO$_6$ (viewed from a axis), (b) IrO$_6$ tetrahedra.}
    \label{fig:crystal_structure}
\end{figure}
\ce{Ca4IrO6} $-$ a combination of \ce{CaIrO3} and 3CaO $-$ crystallizes in $R\bar{3}c$ space group \cite{CalderPRB14}. A unit cell, depicted in \cref{fig:crystal_structure}(a), comprises six formula units with nominal oxidation states Ca$_4^{2+}$Ir$^{4+}$O$_6^{2-}$. Each Ir atom is surrounded by six O atoms forming an octahedron, as illustrated in \cref{fig:crystal_structure}(b). In order to simulate electron and hole doping in the system with varying doping concentrations, we consider partial substitution of Ca atoms with La and Na atoms, respectively, leading to a formula Ca$_{4-x}$M$_x$IrO$_6$ of the doped compounds, M representing La or Na, and $x \in \{0.17, 0.33, 0.50\}$. All our calculations of total energy, electronic structure, and magnetic properties are carried out using density functional theory (DFT), as implemented in the {\scshape vasp} code \cite{vasp1,vasp2}. A plane-wave basis set subject to a $500$ eV kinetic energy cutoff along with projector augmented wave (PAW) method \cite{paw} is used for describing the potential. The Brillouin zone integration has been performed using the improved tetrahedron method \cite{BlochlPRB94T} over a $5 \times 5 \times 5$ $\Gamma$-centered $k$-point mesh. The exchange-correlation functional in DFT is approximated through local (spin) density approximation (L(S)DA) \cite{ldaCA,PerdewPRB81} with a Hubbard-$U$ correction for electron-electron correlation of Ir-$5d$ states, the so-called LDA+$U$ method \cite{DudarevPRB98}, setting $U - J = 1.5$~eV. The atomic positions and the lattice vectors are optimized to minimize the Hellman-Feynman force on each atom and the stress on the simulation cell, respectively, to a tolerance of $10^{-2}$~eV~\AA$^{-1}$. The structures of pristine Ca$_4$IrO$_6$ and the doped systems have been separately optimized to account for local structural changes upon doping. Spin-orbit interaction has not been considered for structural optimization to avoid too expensive computation for almost no change in the structure.
\begin{figure*}
	\includegraphics[scale=0.85]{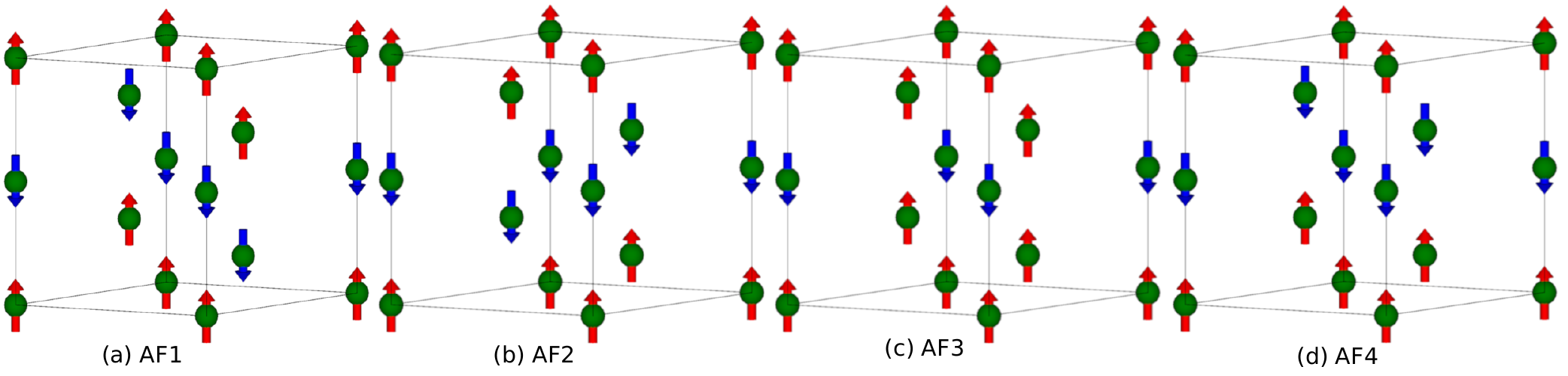}
	\caption{\label{fig:IniMagConfig} From left to right all four different anti-ferromagnetic configurations.}
\end{figure*}
The symmetric, asymmetric, and anisotropic magnetic interactions may be described by the Hamiltonian \cite{ChakrabortyPRB18}
\begin{equation}
    H = - \sum_{i,j} J_{ij} \vec{S}_i \cdot \vec{S}_j + \sum_{\langle i,j \rangle} \vec{D}_{ij} \cdot (\vec{S}_i \times \vec{S}_j) + \sum_i \epsilon_\text{an}^i |\vec{S}_i|^2, \label{eq:magH}
\end{equation}
where $J_{ij}$ represents the coefficient of symmetric (Heisenberg) interaction between the $i^\text{th}$ and the $j^\text{th}$ spins, $\vec{S}_i, \vec{S}_j$ are the corresponding spin vectors, $\vec{D}_{ij}$ is the vector coefficient to asymmetric (Dzyaloshinskii-Moriya) interaction, and $\epsilon_\text{an}^i$ is the anisotropy coefficient for the $i^\text{th}$ spin. Besides ferromagnetic arrangement, we have considered few possible antiferromagnetic arrangements of the spins on Ir atoms that manifest themselves in a noncollinear fashion upon convergence, as depicted in \cref{fig:IniMagConfig}, denoted by AF1, AF2, AF3, and AF4 arrangements. While AF3 configuration, having a nonzero total magnetic moment, is not an antiferromagnetic arrangement in a strict sense, we consider this configuration to assess if it is energetically more favorable. The $J$-values for symmetric interaction in \cref{eq:magH} may be calculated for the system using the following equations involving the total energies of the magnetic arrangements:
\begin{figure}
    \centering
    \includegraphics[scale=0.65]{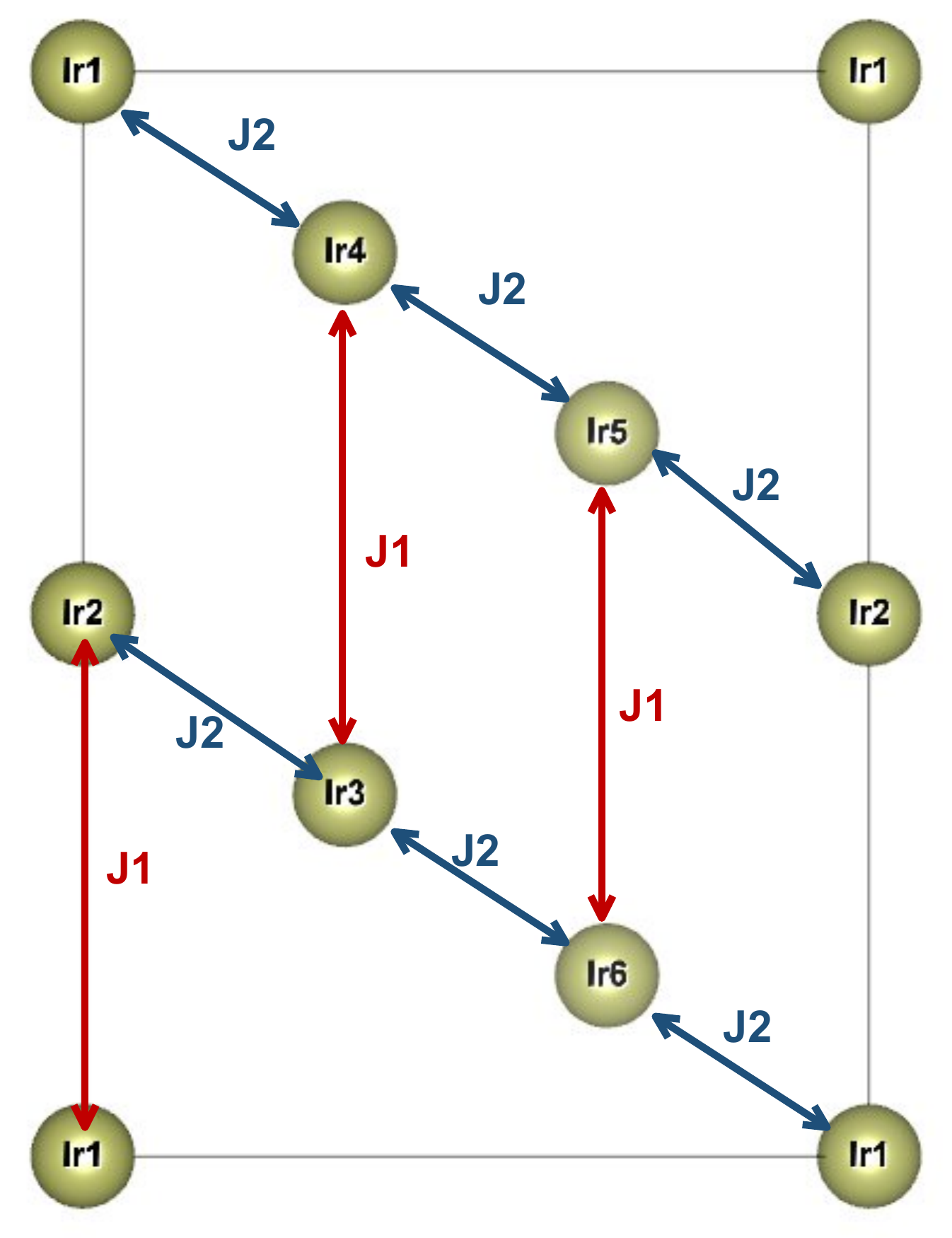}
    \caption{\label{fig:exchangeInteraction}The magnetic exchange paths $J_1$ and $J_2$ are depicted here.}
    
\end{figure}
\begin{align}
    \varepsilon_\text{FM} &= -(3J_1 + 6J_2) \frac{N^2}{4} \nonumber \\
    \varepsilon_\text{AF1} &= -(-J_1 - 2J_2) \frac{N^2}{4} \nonumber \\
    \varepsilon_\text{AF2} &= -(-3J_1 - 6J_2) \frac{N^2}{4} \label{eq:J} \\
    \varepsilon_\text{AF3} &= -(J_1 + 2J_2) \frac{N^2}{4} \nonumber \\
    \varepsilon_\text{AF4} &= -(J_1 - 2J_2) \frac{N^2}{4} \nonumber
\end{align}
where, $N$ represents the number unpaired electrons at Ir-sites. \cref{fig:exchangeInteraction} depicts the exchange paths $J_1$ and $J_2$ between Ir-sites. The Dzyaloshinskii-Moriya (DM) interaction coefficients are calculated by localizing the magnetic moments of two Ir sites, as described in refs. \cite{ChakrabortyPRB18,YangPRL15}.
\section{\label{sec:result}Results and Discussions}
We systematically discuss our results for pristine \ce{Ca4IrO6} and the electron and hole-doped systems at three different doping concentrations.
\subsection{Pristine \ce{Ca4IrO6}}
\begin{figure}
	\centering
	\includegraphics[scale=0.36]{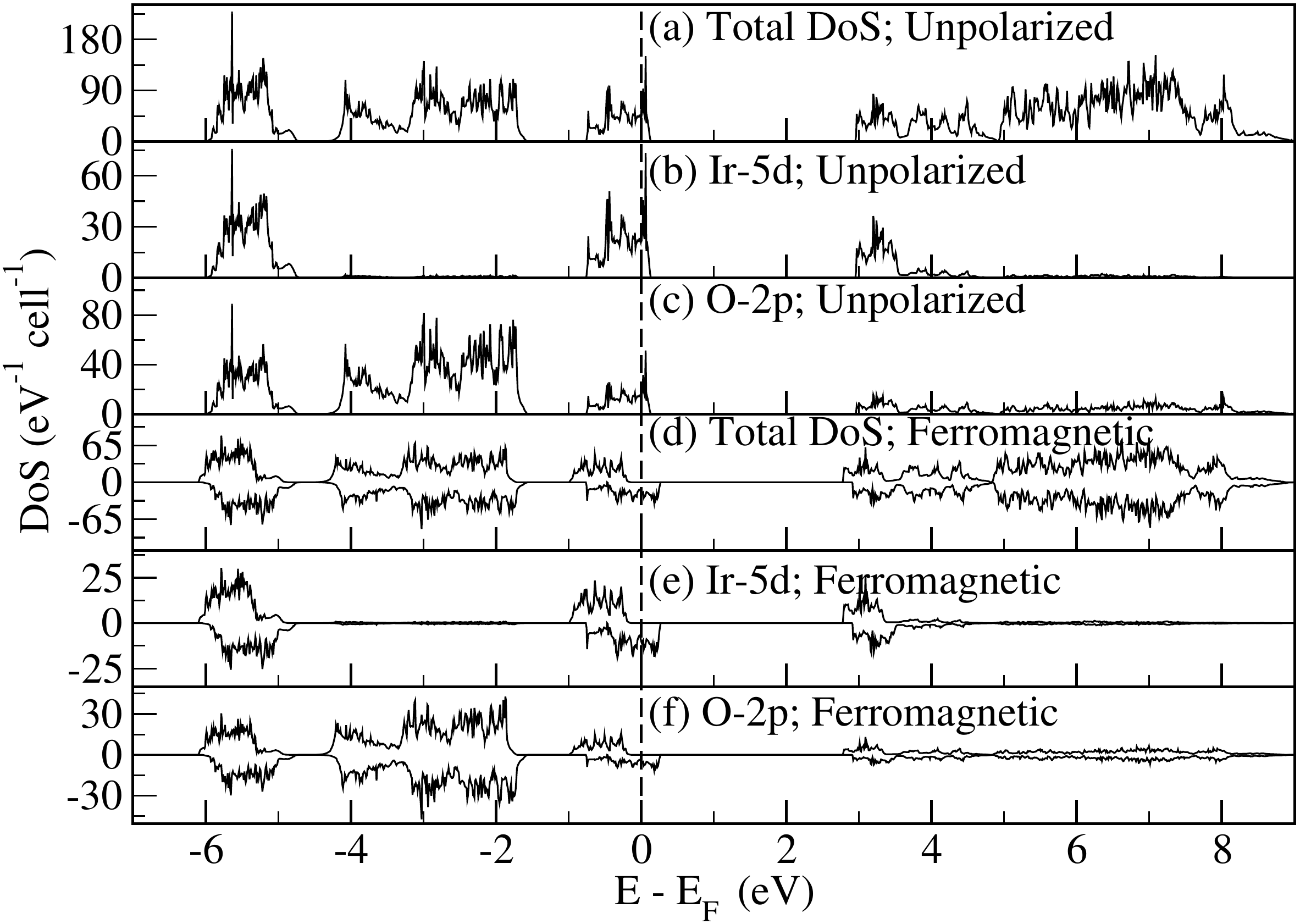}
	\caption{\label{fig:DOS-NSP-FM_PC}The total spin-unpolarized density of states for \ce{Ca4IrO6} is shown in (a), while (b) and (c) displays the orbital-projected density of states for Ir-$5d$ and O-$2p$ states respectively. Upon spin-polarization and considering ferromagnetic arrangement of the spins, the total and Ir-$5d$, O-$2p$ orbital projected density of states are shown in (d), (e), and (f), respectively.}
\end{figure}
In order to understand the implications of electron and hole-doping in \ce{Ca4IrO6}, we first carefully study the electronic structure of pristine \ce{Ca4IrO6}. The density of states (DoS) for spin-unpolarized and spin-polarized calculations has been shown in \cref{fig:DOS-NSP-FM_PC}, suggesting a conducting state within our LDA+$U$ calculations when spin-orbit interaction has not been considered, contrasting the previous report of a Mott-insulator ground state \cite{CalderPRB14}. Comparing the spin-unpolarized total DoS shown in \cref{fig:DOS-NSP-FM_PC}(a) with the orbital projected DoS in \cref{fig:DOS-NSP-FM_PC}(b) and \cref{fig:DOS-NSP-FM_PC}(c), we gather that a hybridized Ir-$5d$ and O-$2p$ orbitals constitute the electronic states near the Fermi level. Besides minor hybridization with Ir-$5d$ orbitals, O-$2p$ orbitals are completely occupied and lie far below the Fermi level, satisfying a nominal O$^{2-}$ oxidation state. We note that while Ir-$5d$ orbitals are partially filled, Ca-$4s$ orbitals are empty and lie far above the Fermi level. Upon spin-polarization, the total DoS and the Ir-$5d$ projected DoS for ferromagnetic arrangement displayed in \cref{fig:DOS-NSP-FM_PC}(d) and \cref{fig:DOS-NSP-FM_PC}(e), respectively, exhibit a relative shift in energy for the majority and minority spin states near the Fermi level, characteristic of net magnetization in the system.
\begin{table*}
    \centering
    \caption{\label{tab:Energy_magmom_LDA+U}The energy $\varepsilon$ relative to the magnetic arrangement with lowest energy and the site-projected magnetic moments $m$ at Ir and O sites are tabulated for \ce{Ca4IrO6}, Ca$_{3.5}$La$_{0.5}$IrO$_6$, and Ca$_{3.5}$Na$_{0.5}$IrO$_6$ without considering spin-orbit interaction.}
    \begin{ruledtabular}
    \begin{tabular}{l|...|...|...}
         & \multicolumn{3}{c|}{\ce{Ca4IrO6}} & \multicolumn{3}{c|}{Ca$_{3.5}$La$_{0.5}$IrO$_6$} & \multicolumn{3}{c}{Ca$_{3.5}$Na$_{0.5}$IrO$_6$} \\
        \cline{2-10}
        & \multicolumn{1}{c}{$\varepsilon$ (meV)} & \multicolumn{1}{c}{$m_\text{Ir}$ ($\mu_B$)} & \multicolumn{1}{c|}{$m_\text{O}$ ($\mu_B$)} & \multicolumn{1}{c}{$\varepsilon$ (meV)} & \multicolumn{1}{c}{$m_\text{Ir}$ ($\mu_B$)} & \multicolumn{1}{c|}{$m_\text{O}$ ($\mu_B$)} & \multicolumn{1}{c}{$\varepsilon$ (meV)} & \multicolumn{1}{c}{$m_\text{Ir}$ ($\mu_B$)} & \multicolumn{1}{c}{$m_\text{O}$ ($\mu_B$)} \\
        \hline
        FM & 0.00 & 0.60 & 0.06 & 6.44 & 0.51 & 0.05 & 26.30 & 0.91 & 0.10 \\
        AF1 & 282.38 & 0.45 & 0.04 & 0 & 0.49 & 0.05 & 36.86 & 0.90 & 0.10 \\
        AF2  &  104.61 & 0.55 & 0.05 &  6.42 &  0.51 & 0.05 &  0     &  0.87 & 0.10 \\
        AF3  &  - & - & - &  6.42 &  0.51 & 0.06 &  44.40 &  0.93 & 0.13 \\
        AF4  &  104.61 & 0.55 & 0.05 &  6.42 &  0.51 & 0.05 &  40.00 &  0.95 & 0.14 \\
    \end{tabular}
    \end{ruledtabular}
\end{table*}
A comparison of the total energies of various magnetic configurations and the site-projected magnetic moments presented at \cref{tab:Energy_magmom_LDA+U} reveals that the ferromagnetic configuration for pristine \ce{Ca4IrO6} has the lowest energy, with a notable projected magnetic moment of 0.60~$\mu_B$ per Ir atom and a minor spillover magnetic moment of 0.06~$\mu_B$ per O atom when spin-orbit interaction has not been considered. We note that our calculations did not converge in the AF3 configuration for the parent compound in the absence of spin-orbit interaction; therefore, no result could be reported. The magnetic exchange interaction strengths along different interaction paths $J_1$ and $J_2$ depicted in \cref{fig:exchangeInteraction} are calculated and tabulated in \cref{tab:exchange}, corroborating the observed lowest energy ferromagnetic state.
\begin{table}
\caption{\label{tab:exchange}The exchange interaction strengths in meV between the Ir atoms along $J_1$ and $J_2$ exchange paths marked in \cref{fig:exchangeInteraction} without considering spin-orbit interaction.}
\begin{ruledtabular}
\begin{tabular}{c....}
Path & \multicolumn{1}{c}{Distance (\AA)} & \multicolumn{1}{c}{\ce{Ca4IrO6}} & \multicolumn{1}{c}{Ca$_{3.5}$La$_{0.5}$IrO$_6$} & \multicolumn{1}{c}{Ca$_{3.5}$Na$_{0.5}$IrO$_6$} \\
\hline
$J_1$ & 5.63 & 9.48 & -40.16 & -4.38 \\
$J_2$ & 5.75 & 1.14 & 20.0 & 1.97 \\
\end{tabular}
\end{ruledtabular}
\end{table}

Subsequently, we incorporate spin-orbit interaction in our calculations and analyze the electronic structure of \ce{Ca4IrO6}.
\begin{figure}
	\includegraphics[scale=1.0]{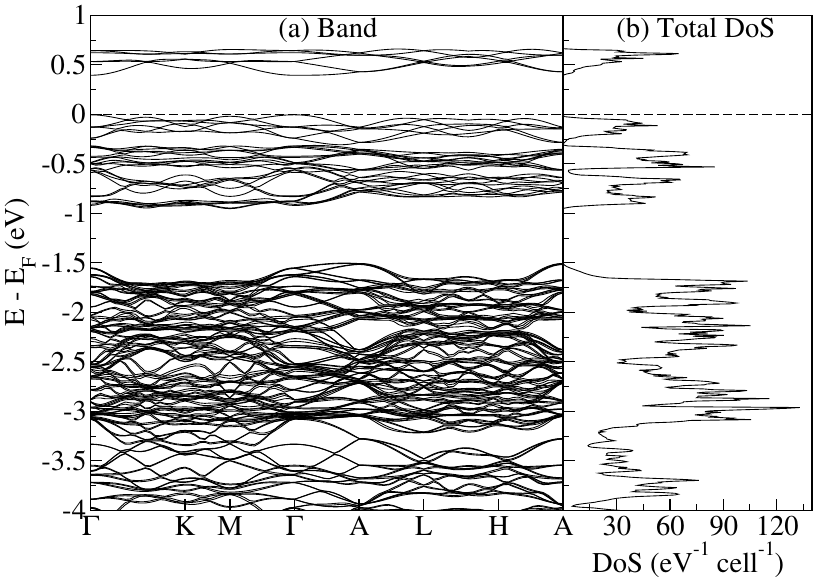}
	\caption{\label{fig:SOC-AFM_PC}(a) Band structure, (b)Total DoS, for Ca$_4$IrO$_6$ with LDA+SOC+U.}
\end{figure}
The band structure and DoS displayed in \cref{fig:SOC-AFM_PC}(a) and \cref{fig:SOC-AFM_PC}(b), respectively, reveal drastic changes in the electronic structure of the Ir bands, leading to a $J_\text{eff} = 1/2$ Mott-insulating state with a gap of $\sim$0.35~eV, in agreement with ref \cite{CalderPRB14}.
\begin{table*}
    \centering
    \caption{\label{tab:Energy_magmom_LDA+SOC+U}The energy $\varepsilon$ relative to the magnetic arrangement with lowest energy and the site-projected spin (orbital) magnetic moments $m$ at Ir and O sites are tabulated for \ce{Ca4IrO6}, Ca$_{3.5}$La$_{0.5}$IrO$_6$, and Ca$_{3.5}$Na$_{0.5}$IrO$_6$ considering spin-orbit interaction for the calculations.}
    \begin{ruledtabular}
    \begin{tabular}{l|...|...|...}
         & \multicolumn{3}{c|}{\ce{Ca4IrO6}} & \multicolumn{3}{c|}{Ca$_{3.5}$La$_{0.5}$IrO$_6$} & \multicolumn{3}{c}{Ca$_{3.5}$Na$_{0.5}$IrO$_6$} \\
        \cline{2-10}
        & \multicolumn{1}{c}{$\varepsilon$ (meV)} & \multicolumn{1}{c}{$m_\text{Ir}$ ($\mu_B$)} & \multicolumn{1}{c|}{$m_\text{O}$ ($\mu_B$)} & \multicolumn{1}{c}{$\varepsilon$ (meV)} & \multicolumn{1}{c}{$m_\text{Ir}$ ($\mu_B$)} & \multicolumn{1}{c|}{$m_\text{O}$ ($\mu_B$)} & \multicolumn{1}{c}{$\varepsilon$ (meV)} & \multicolumn{1}{c}{$m_\text{Ir}$ ($\mu_B$)} & \multicolumn{1}{c}{$m_\text{O}$ ($\mu_B$)} \\
        \hline
        FM & 1.60 & 0.12 (0.32) & 0.02 & 17.02 & 0.13 (0.30) & 0.02 & - & - & - \\
        AF1 & 0 & 0.11 (0.32) & 0.03 & 17.13 & 0.13 (0.28) & 0.02 & 2.24 & 0.42 (0.44) & 0.03 \\
        AF2 & 1.46 & 0.12 (0.32) & 0.03 & 0 & 0.14 (0.27) & 0.02 & 0 &  0.21 (0.32) & 0.02 \\
        AF3  &  1.26 & 0.12 (0.32) & 0.03 &  17.02 &  0.14 (0.28) & 0.02 & 0.24 &  0.12 (0.28) & 0.02 \\
        AF4  &  1.46 & 0.12 (0.32) & 0.03 &  17.82 &  0.27 (0.33) & 0.03 & 0.72 &  0.20 (0.28) & 0.02 \\
    \end{tabular}
    \end{ruledtabular}
\end{table*}
Upon considering spin-orbit interaction in our calculations, the relative energies of various magnetic configurations and the corresponding projected spin (orbital) magnetic moments have been tabulated in \cref{tab:Energy_magmom_LDA+SOC+U}. This table shows that the AF1 antiferromagnetic state becomes the lowest energy magnetic configuration, and the differences in energy among the magnetic configurations are substantially reduced. Further, the projected spin moments of Ir atoms become smaller, with a significant orbital moment. We also notice a canted arrangement of the magnetic moments, as depicted in \cref{fig:magconfig}(a). Our calculations did not reveal any significant DM interaction operating in the system. The spin-orbit interaction leads to substantially strong magneto-crystalline anisotropy in the system, with an energy difference of 5.9~meV between spin-quantization axes (100) and (001), (001) being the easy axis. In this context, a recent report experimentally demonstrates that nonmagnetic impurity may lead to noncollinear magnetic order in the presence of magnetic frustration or anisotropy in the system, hinting at a substantially noncollinear magnetic arrangement in the doped compounds \cite{ParkNC21}.

\subsection{Electron and hole doping}
\begin{figure}
	\includegraphics[scale=1.75]{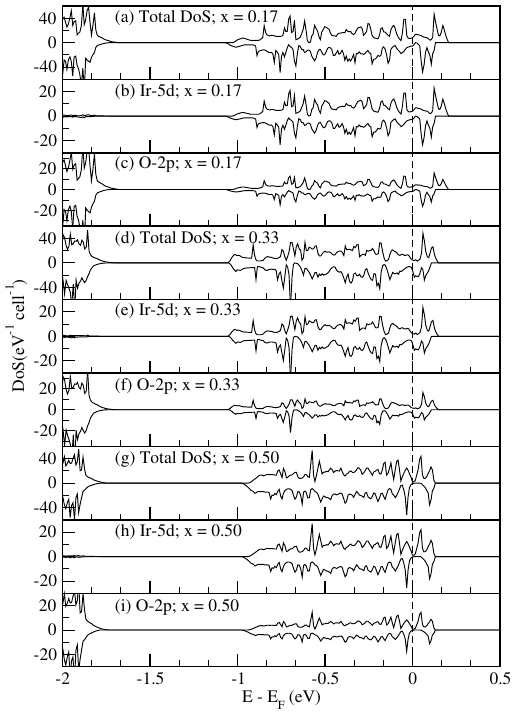}
	\caption{\label{fig:DOS-AFM_La}The total density of states along with Ir-$5d$ and O-$2p$ orbital projected density of states for Ca$_{4-x}$La$_x$IrO$_{6}$ is shown here with $x \in \{0.17, 0.33, 0.50\}$, spin-orbit interaction has not been considered.}
\end{figure}

\begin{figure}
	\includegraphics[scale=1.85]{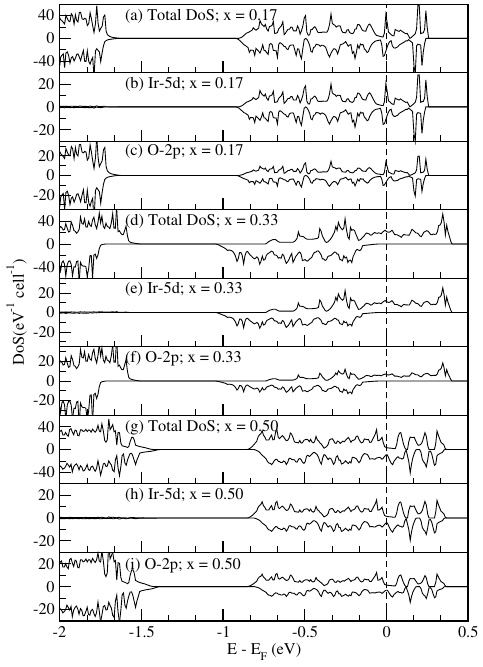}
	\caption{\label{fig:DOS-AFM_Na}The total density of states along with Ir-$5d$ and O-$2p$ orbital projected density of states for Ca$_{4-x}$Na$_x$IrO$_{6}$ is shown here with $x \in \{0.17, 0.33, 0.50\}$, spin-orbit interaction has not been considered.}
\end{figure}

After understanding the electronic structure and magnetic properties of \ce{Ca4IrO6}, we analyze the consequences of doping the system with electrons and holes by partially substituting Ca atoms with La or Na, respectively, leading to the general formula Ca$_{4-x}$M$_x$IrO$_6$ of the doped compounds, M representing La or Na with $x \in \{0.17, 0.33, 0.50\}$. \Cref{tab:Energy_magmom_LDA+U} shows the relative energies in different magnetic configurations, revealing AF1 and AF2 to be the lowest-energy configurations for Ca$_{4-x}$La$_x$IrO$_{6}$ and Ca$_{4-x}$Na$_x$IrO$_{6}$, respectively. While doping La (electron doping) in place of Ca does not significantly influence the projected magnetic moment on Ir, doping Na (hole doping) almost doubles it. We attribute the introduction of vacancy in the covalent bond due to hole doping, thereby reducing the covalent character of the bond to the increment of magnetic moments at the Ir sites. The density of states presented in \cref{fig:DOS-AFM_La} and \cref{fig:DOS-AFM_Na} for Ca$_{4-x}$La$_x$IrO$_{6}$ and Ca$_{4-x}$Na$_x$IrO$_{6}$ reveals progressive filling and emptying of the Ir-$5d$ states, respectively, with increasing doping concentration.
\begin{figure*}
	\includegraphics[scale=0.38]{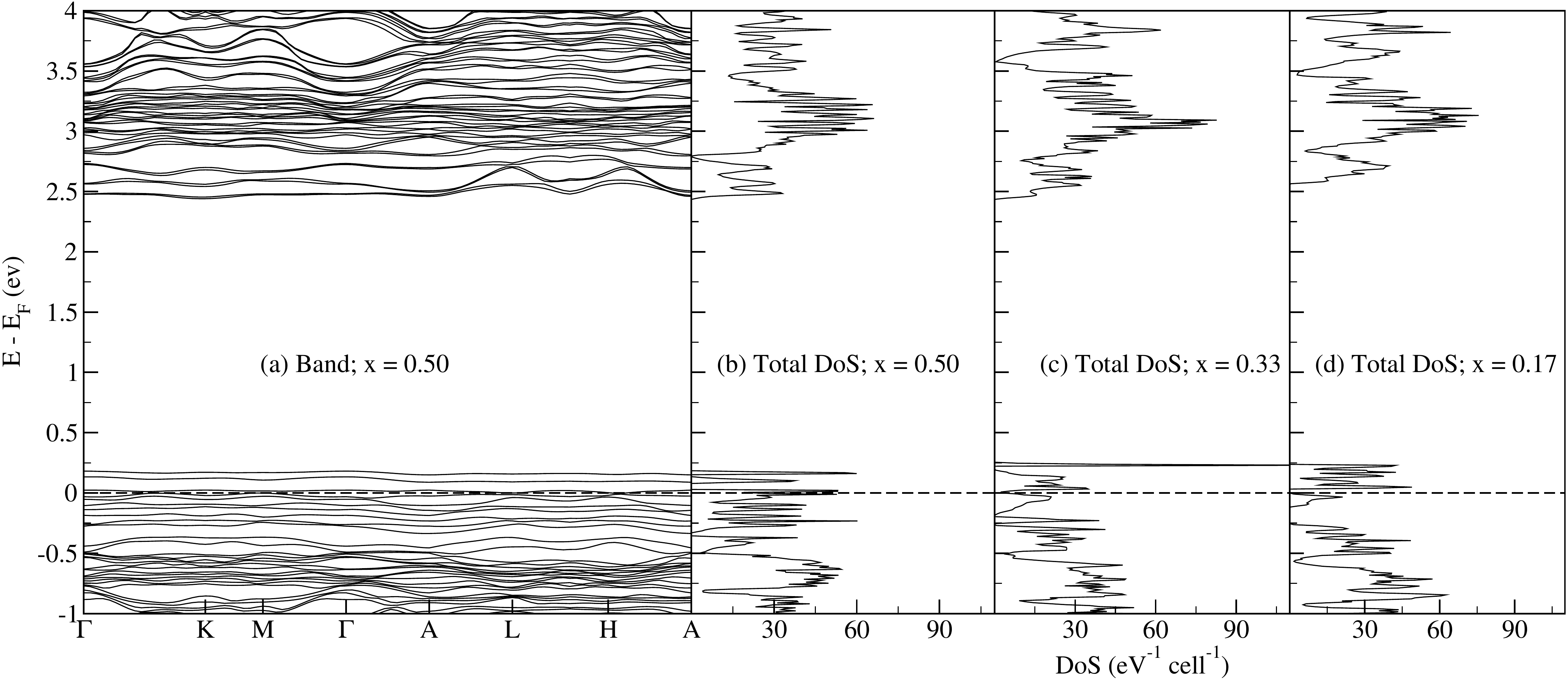}
	\caption{\label{fig:SOC-AFM_La}The electronic structure of Ca$_{4-x}$La$_x$IrO$_{6}$ calculated considering spin-orbit interaction is shown here. (a) displays the band structure for $x = 0.50$, while (b), (c), and (d) represent the DoS for $x = 0.50$, 0.33, and 0.17, respectively.}
\end{figure*}
\begin{figure*}
	\includegraphics[scale=1.16]{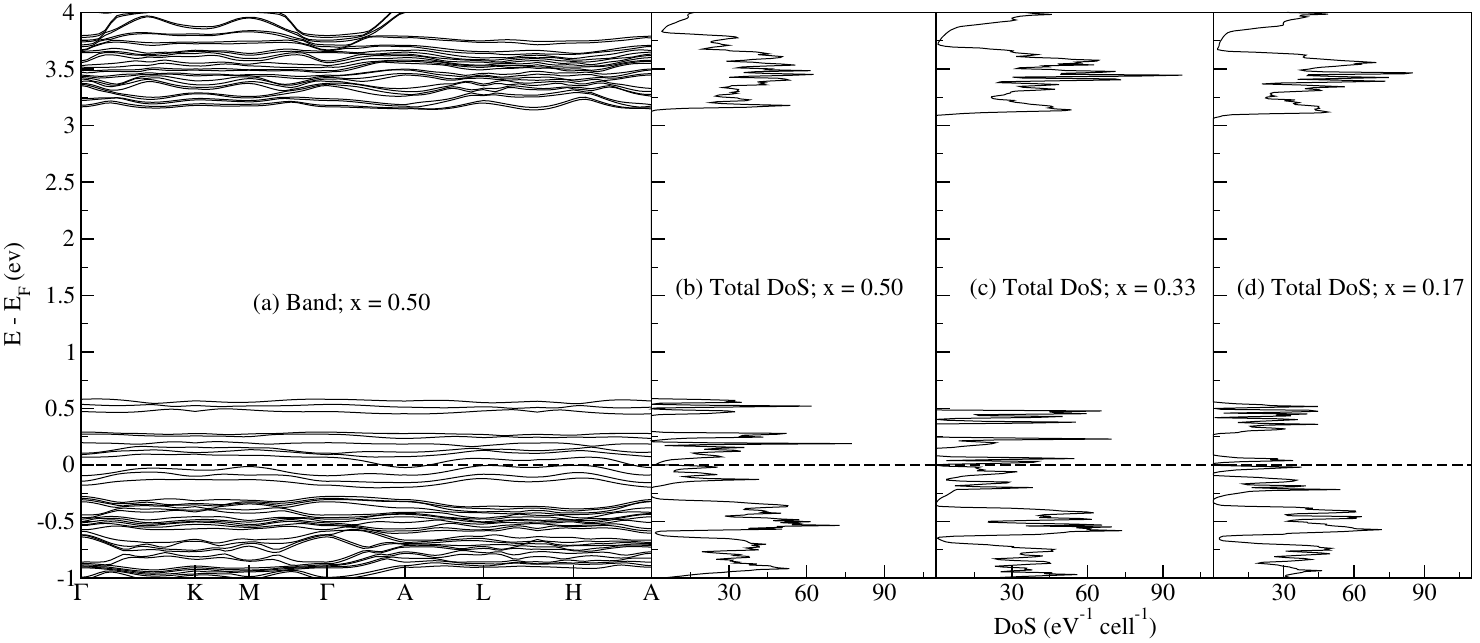}
	\caption{\label{fig:SOC-AFM_Na}The electronic structure of Ca$_{4-x}$Na$_x$IrO$_{6}$ calculated considering spin-orbit interaction is shown here. (a) displays the band structure for $x = 0.50$, while (b), (c), and (d) represent the DoS for $x = 0.50$, 0.33, and 0.17, respectively.}
\end{figure*}
\begin{figure*}
	\includegraphics[scale=1.1]{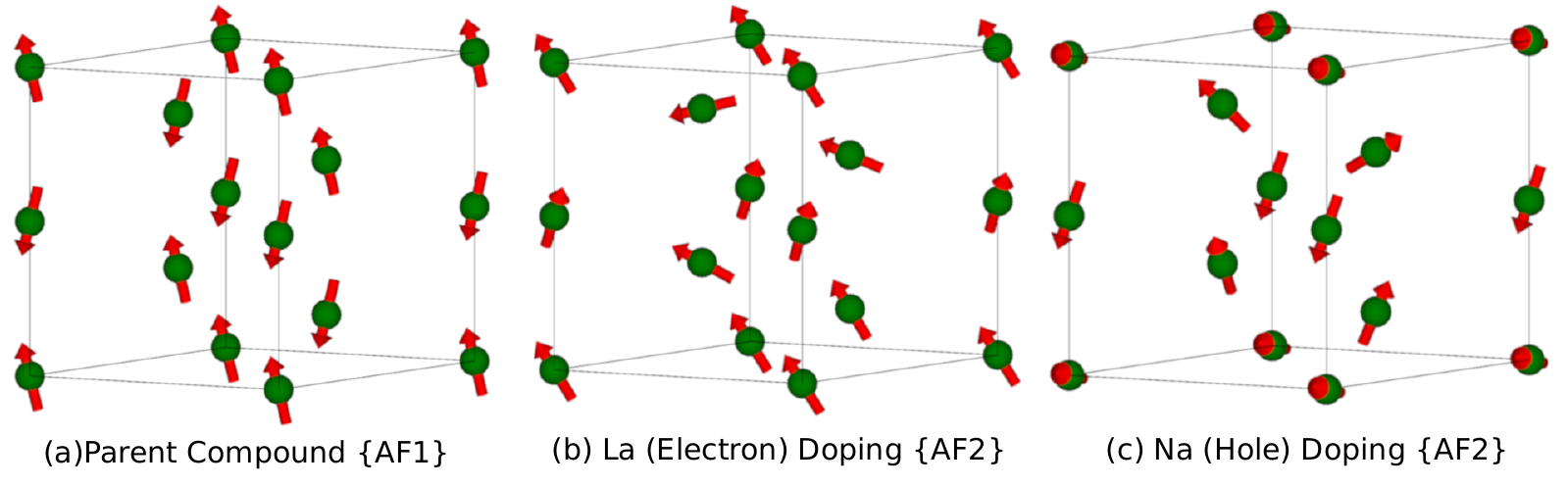}
	\caption{\label{fig:magconfig} (a) Ground state magnetic configuration of \ce{Ca4IrO6}, (b) Ground state magnetic configuration of Ca$_{3.5}$La$_{0.5}$IrO$_6$, and (c) Ground state magnetic configuration of Ca$_{3.5}$Na$_{0.5}$IrO$_6$ considering spin-orbit interaction}
\end{figure*}
We note that the lowest energy magnetic configuration for the Na-doped system with $x = 0.33$ retains some magnetic moment as the arrangement of spins does not lead to a complete cancellation. \Cref{fig:DOS-AFM_Na}(d)(e)(f) exhibit an energy shift in the DoS between the majority and the minority spin states, corroborating the residual magnetic moment. The magnetic exchange interaction strengths along the paths $J_1$ and $J_2$ for the electron- and hole-doped compounds have been tabulated in \cref{tab:exchange}, justifying a stable antiferromagnetic state in the doped compounds.

Considering spin-orbit interaction in our calculations, first, we estimated the relative energies in the ferromagnetic and various antiferromagnetic configurations for La- and Na-doped systems, as tabulated in \cref{tab:Energy_magmom_LDA+SOC+U}. Our results indicate AF2 configuration has the lowest energy for the doped compounds, as opposed to AF1 having the lowest energy in pristine \ce{Ca4IrO6}. The projected spin and orbital magnetic moments on Ir atoms for the doped compounds remain in a similar ballpark range as the parent compound. We plotted the band dispersion and density of states, as displayed in \cref{fig:SOC-AFM_La} and \cref{fig:SOC-AFM_Na} for La-doped and Na-doped systems, respectively. We gather from these figures that upon doping of La (electron-doping), the bands above the Mott-like gap get progressively filled with increasing doping concentration. In contrast, upon doping Na (hole-doping), the bands right below the gap become partially empty. Further, the Mott-like gap near the Fermi level shrinks with increasing electron-doping concentration while remaining almost unaltered in case of hole-doping. While $J_\text{eff} = 1/2$ insulating state is no longer preserved upon doping, the overall antiferromagnetic nature of the compound is still retained, albeit with a drastic change in the microscopic magnetic properties. As seen from \cref{fig:magconfig}, the canted magnetic moments in pristine \ce{Ca4IrO6} transform into highly a noncollinear arrangement for both electron and hole doping. However, our calculations reveal no significant DM interaction in the system. Therefore, the highly noncollinear arrangement of spins can be attributed to strong magnetocrystalline anisotropy in \ce{Ca4IrO6} that becomes even stronger in the doped compounds. Our calculations reveal anisotropy energies of 14.0~meV and 13.3~meV for the La-doped, and Na-doped compounds, respectively; when calculated along (001) and (100) directions, (001) direction is preferred. We note that doping of nonmagnetic La and Na ions substantially increases the anisotropy from its value of 5.9~meV in the parent compound. Our results are consistent with the recent experimental observation that nonmagnetic impurities can transform a frustrated or anisotropic antiferromagnet into a highly noncollinear antiferromagnet \cite{ParkNC21}.

\section{\label{sec:conc}Conclusion}
We have investigated the electronic structure and magnetic properties of a $J_\text{eff} = 1/2$ iridate \ce{Ca4IrO6} and electron and hole doping of the same using first-principles DFT calculations. Our results considering spin-orbit interaction reveal a canted antiferromagnetic Mott-insulator ground state for the parent compound that transforms into a highly noncollinear antiferromagnetic conductor upon doping of nonmagnetic impurities. Increasing electron doping concentration by substituting Ca with La (hole doping by replacing Ca with Na) has resulted in the progressive filling of bands above the Fermi level (progressive emptying of bands below the Fermi level), leading to a conducting state. We find negligible Dzyaloshinskii–Moriya interaction strength in the pristine and the doped compounds, suggesting the highly noncollinear nature of magnetic moments may be attributed to magnetic frustrations arising due to relatively strong spin-orbit interaction and magneto-crystalline anisotropy in the system. Our results may be necessary for understanding magnetic properties and spin textures of conducting iridates and other conducting materials with strong spin-orbit interaction, therefore interesting from the viewpoint of spintronics.
\begin{acknowledgments}
Financial support from SERB India through grant numbers ECR/2016/001004, CRG/2021/005320, DST India INSPIRE fellowship through grant number IF171000, and the use of the high-performance computing facility at IISER Bhopal are gratefully acknowledged.
\end{acknowledgments}

%
\end{document}